\documentclass[prb,twocolumn,superscriptaddress,showpacs,%
preprintnumbers,amsmath,amssymb]{revtex4}
\usepackage{amsmath}
\usepackage{graphicx}
\usepackage{dcolumn}

\begin{document}
\title{Structural and dynamical properties of a $quasi$-one-dimensional classical binary system}
%%%%%%%%%%%%%%%%%%%%%%%%%%%%%%%%%%%%%%%%%%%%%%%%%%%%%%%%%%%%%%%%%%%%%%%%%%%%%%%%%%%%%%%%%%%%%%%%%%%%%%%%%%%%%%
\author{W.~P.~Ferreira}
\email{wandemberg@fisica.ufc.br} \affiliation{Departamento de F\'isica,
Universidade Federal do Cear\'a, Caixa Postal 6030, Campus do Pici, 60455-760 Fortaleza, Cear\'a, Brazil}%
%%%%%%%%%%%%%%%%%%%%%%%%%%%%%%%%%%%%%%%%%%%%%%%%%%%%%%%%%%%%%%%%%%%%%%%%%%%%%%%%%%%%%%%%%%%%%%%%%%%%%%%%%%%%%%
\author{J.~C.~N.~Carvalho}
\affiliation{Departamento de F\'isica,
Universidade Federal do Cear\'a, Caixa Postal 6030, Campus do Pici, 60455-760 Fortaleza, Cear\'a, Brazil}%
%%%%%%%%%%%%%%%%%%%%%%%%%%%%%%%%%%%%%%%%%%%%%%%%%%%%%%%%%%%%%%%%%%%%%%%%%%%%%%%%%%%%%%%%%%%%%%%%%%%%%%%%%%%%%%
\author{P.~W.~S.~Oliveira}
%\email{pwfisica@fisica.ufc.br}
\affiliation{Departamento de F\'isica,
Universidade Federal do Cear\'a, Caixa Postal 6030, Campus do Pici, 60455-760 Fortaleza, Cear\'a, Brazil}%
%%%%%%%%%%%%%%%%%%%%%%%%%%%%%%%%%%%%%%%%%%%%%%%%%%%%%%%%%%%%%%%%%%%%%%%%%%%%%%%%%%%%%%%%%%%%%%%%%%%%%%%%%%%%%%
\author{G.~A.~Farias}
\affiliation{Departamento de F\'isica,
Universidade Federal do Cear\'a, Caixa Postal 6030, Campus do Pici, 60455-760 Fortaleza, Cear\'a, Brazil}%
%%%%%%%%%%%%%%%%%%%%%%%%%%%%%%%%%%%%%%%%%%%%%%%%%%%%%%%%%%%%%%%%%%%%%%%%%%%%%%%%%%%%%%%%%%%%%%%%%%%%%%%%%%%%%%
\author{F.~M.~Peeters}
\email{francois.peeters@ua.ac.be} \affiliation{Department of Physics, University of Antwerp,
Groenenborgerlaan 171, B-2020 Antwerpen, Belgium}
%%%%%%%%%%%%%%%%%%%%%%%%%%%%%%%%%%%%%%%%%%%%%%%%%%%%%%%%%%%%%%%%%%%%%%%%%%%%%%%%%%%%%%%%%%%%%%%%%%%%%%%%%%%%%%
\date{ \today }

\begin{abstract}

The ground state configurations and the \lq{}\lq{}normal\rq{}\rq{} mode spectra of a $quasi$-one-dimensional (Q1D)
binary system of charged particles interacting through a screened Coulomb potential are presented.
The minimum energy configurations were obtained analytically and independently through molecular
dynamic simulations. A rich variety of ordered structures were found as a function of the
screening parameter, the particle density, and the ratio between the charges of the distinct types
of particles. Continuous and discontinuous structural transitions, as well as an unexpected
symmetry breaking in the charge distribution are observed when the density of the system is
changed. For near equal charges we found a disordered phase where a mixing of the two types of
particles occurs. The phonon dispersion curves were calculated within the harmonic approximation
for the one- and two-chain structures.

\end{abstract}


\maketitle

\section{Introduction}
\label{sec:introduction}

Advances in experimental and numerical techniques allowed a considerable improvement of the
possibilities to create and study new crystalline structures which will open a vast field of
scientific and technological applications. Recently, a variety of novel three-dimensional (3D) binary super-lattices
obtained through the combination of semiconducting, metallic, and magnetic colloidal
nano-particles was observed experimentally\cite{leunissen05,shevchenko06}. The self-organization
in a two-dimensional (2D) binary colloidal system resulted in interesting mixed configurations
with partial clustering \cite{hoffmann06}. The use of colloidal particles is technically
interesting since it allows a real time and spatial direct observation of their position which
is a great advantage as compared to atoms or molecules,
as shown recently in an interesting experimental study concerning defect induced
melting\cite{alsayed05}.

Systems of particles moving in space of a reduced dimensionality or submitted to an external
confinement exhibit different behavior from their free-of-border counterparts. It has been shown
previously that the presence of constraints generates new features which are in some cases
unexpected, as e.g. the curious ground state configuration found in a 3D
finite size binary Coulomb cluster\cite{matthey03}, where a mixed arrangement of the two types of
particles was observed in the core, while the edge of the system was characterized by shell
structures of separated species. Similarly, in 2D systems we can cite a reentrant melting
phenomenon in a 2D circular confined cluster of repulsive dipoles \cite{bubeck98,ischweigert00},
and interesting asymmetrical ground state configurations observed in symmetrically confined
Coulomb clusters \cite{wand05,wand07}.

Under specific conditions of density and temperature an electron gas crystallizes in an ordered
structure forming the so called Wigner crystal (WC) \cite{wigner34}. In general, a Wigner crystal
structure can also refer to an ordered phase occurring in other systems of interacting particles.
Several non-electronic systems show such a Wigner ordered phase, which has been experimentally
observed in colloidal systems \cite{liu95,zahn99,golosovsky02}, dusty plasmas
\cite{chu1994,liu03,liu05}, and ion traps \cite{levi1988, drewsen98}.

Depending on the dimensionality and extent of the system the ordered phase can have different
symmetry. As shown theoretically, the minimum energy arrangement of the 2D infinite electron gas
is the triangular lattice \cite{maradudin77}, while the circular confined analog cluster was
observed experimentally \cite{melzer03} and numerically \cite{bedanov94}, and has a mixed ground
state configuration with a triangular structure in the core and a circular structure at the edge
of the system as a consequence of the symmetry of the external confinement
\cite{melzer03,bedanov94}.

When the 2D infinite system is subjected to an extra confinement (e.g. parabolic) in one direction
the system can be called $quasi$-one-dimensional (Q1D). Such a Q1D system of charged particles
interacting through a repulsive potential self-organizes in a chain-like structure were recently
experimentally studied in dusty plasmas \cite{liu03,liu05} and paramagnetic colloids
\cite{haghgooie06,koppl06}, and through analytical and numerical calculations
\cite{haghgooie04,gio04}. In Ref. \onlinecite{gio04}, the authors presented a systematic and
interesting study of the structural, dynamical, and thermal properties as a function of the linear
density of the system. It is claimed that the 1D Coulomb crystal confined in a storage ring may be
used in atomic clocks or even for quantum computation \cite{birkl92,itano93,cirac95}. An
interesting application for the columnar regular structure of interacting particles was shown in
Ref. \onlinecite{doyle02} where a regular linear array of superparamagnetic colloidal particles
confined in a thin gap was used for the separation of DNA.

Besides the single-component system of interacting particles, which has been widely studied in the
last years, Wigner crystallization may also be observed in multi-component systems of interacting
particles \cite{liu06}. The simplest multi-component case is the binary mixture of two types of
particles which exhibits a richer set of physical properties when compared to the one-component
system. For example, a large number of different equilibrium configurations which are
intrinsically dependent on the relative fraction of the different types of particles was recently
found in a 2D system of dipoles \cite{assoud07}. An interesting analysis of the pair correlation
function decay as a function of the packing fraction in a binary system of hard-spheres was
presented in Ref. \onlinecite{baumgartl07}. The presence of particles with distinct physical
properties (e.g. size, charge, mass) introduces a competition between different scales, which is
the reason for the richer phenomenology in such systems.

In the present paper, we extend the work of Ref. \onlinecite{gio04} to a binary two-dimensional
system of charged particles interacting through a screened Coulomb potential and confined in a
parabolic channel. The particles are allowed to move (without friction) in the $x$-$y$ plane, but
due to the external parabolic confinement in the y-direction, the system acquires a Q1D character,
i. e. the particles are free to move in the $x$-direction and have a constrained motion in the
$y$-direction. Due to the possibility to tune the range of the screened interaction potential in
the present model, the binary system of hard-spheres \cite{baumgartl07} becomes a particular case
of our system, which occurs when the screening length of the interaction potential is very small.
In spite of the specific interaction potential considered in the present paper, the main
qualitative features of the obtained results are expected not to depend on the functional form of
the inter-particle interaction.

We present a systematic study of the structure of the ground state as a function of the linear
density, the ratio between the charges of the distinct types of particles, which characterizes the
binary system, and the parameter which determines the range of the repulsive interaction potential
between the particles. The range of the interaction between the charged particles can be varied in
experimental systems of colloidal particles by changing, for instance, the salt concentration in
the solvent \cite{lowen01}. In the present analysis we limit ourselves to an equal density of the
two types of particles which are assumed to have the same mass in order to reduce the number of parameters.
The normal modes, i.e. the phonons, of the present system are also studied.

The paper is organized as follow. In Sec. II, we describe the model and the procedure used to find
the minimum energy configurations and the frequencies of the normal mode spectra. In Sec. III we
study two different zero temperature phase diagrams. In the first, the ground state configurations
are presented for distinct screening parameter of the interaction potential and densities. In the
second, the dependence of the crystal structures are shown for different values of the ratio
between charges and density. The analysis of the phonon dispersion curves is presented in Sec. IV.
Our conclusions are given in Sec. V.

\begin{figure*}
\begin{center}
\includegraphics[scale=0.6]{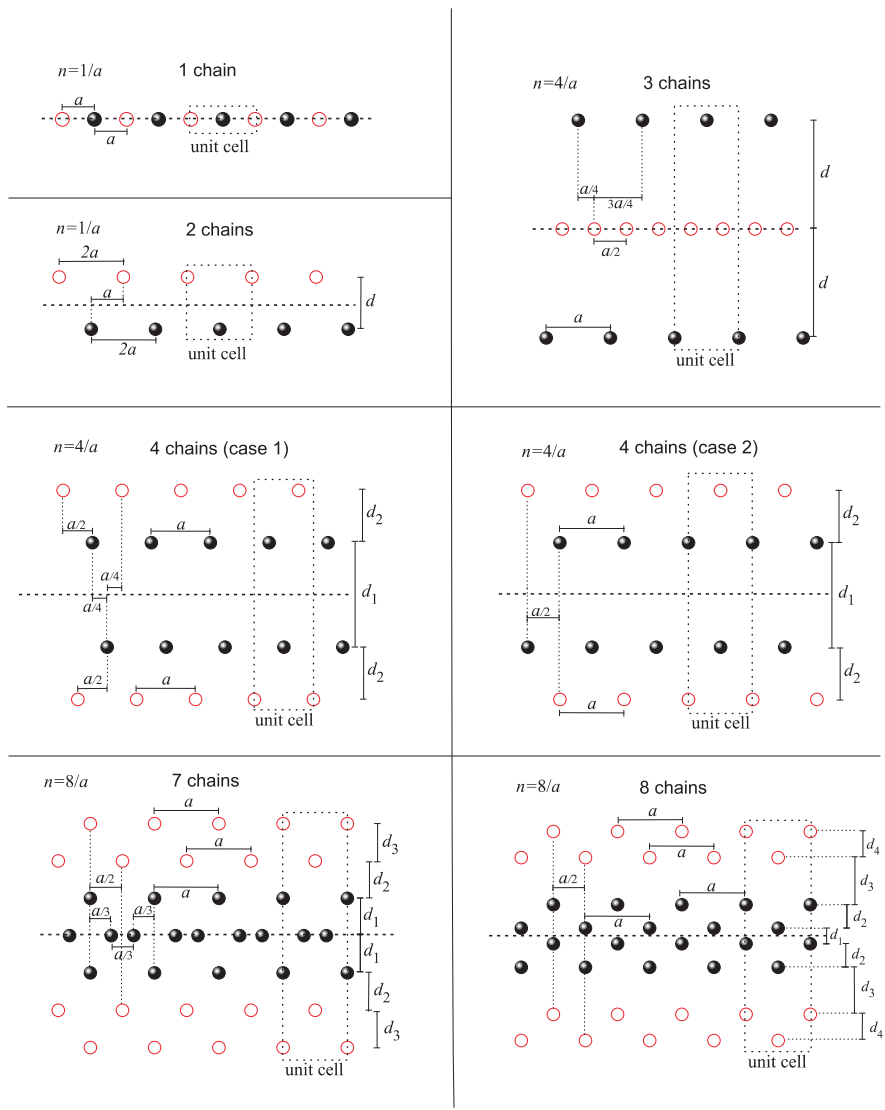}
\caption{(Color online) Ground state configurations. The $y=0$ axis is shown by the thin
horizontal dotted line and the two different types of particles in the binary system are indicated
by different symbols. The red open circles are particles with charge $q_b/q_a=\alpha=2$, and the
black solid spheres are particles with $q_a/e=1$. The linear density of each configuration is also
shown.}\label{fig:configs}
\end{center}
\end{figure*}

\section{Numerical approach}
\label{sec:model}

We studied a two-dimensional binary cluster consisting of a large and equal number of particles
with distinct charges $q_a$ and $q_b$, which are allowed to move in the $x$-$y$ plane. The charged
particles interact through a repulsive Debye-H\"{u}ckel (or Yukawa) potential $exp(-r/\lambda)/r$,
are free to move in the $x$-direction, and are confined by a one-dimensional parabolic potential
which limits the motion of the particles in the $y$-direction. The potential energy of the system
is given by
\begin{equation}\label{hamiltonian}
\begin{split}
  H &= \sum_{i}\frac{1}{2}m \omega_0^2 y_i^2 + \frac{q_{a}q_{b}}{\epsilon}\sum_{m}\sum_{n}
  \frac{exp(- \vert \mathbf{r}_{m} - \mathbf{r}_{n}\vert/\lambda)}{\vert \mathbf{r}_{m} -
  \mathbf{r}_{n}\vert}
   \\& + \frac{q_{a}^{2}}{\epsilon} \sum_{i>j}\frac{exp(-\vert \mathbf{r}_{i} - \mathbf{r}_{j}\vert/\lambda)}{\vert \mathbf{r}_{i} - \mathbf{r}_{j}\vert}
+ \frac{q_b^2}{\epsilon} \sum_{k>l} \frac{exp(-\vert \mathbf{r}_{k} -
\mathbf{r}_{l}\vert)/\lambda}{\vert \mathbf{r}_{k} - \mathbf{r}_{l}\vert},
\end{split}
\end{equation}
where $\epsilon$ is the dielectric constant of the medium the particles are moving in, $\lambda$
is the Debye length, and $r_i \equiv |\mathbf{r}_i|$ is the distance of the $i^{th}$ particle from
the center of the confinement potential. In order to reveal the important parameters of the
system, it is convenient to write the energy and the distances in units of $E_{0}=(m \omega
_{0}^{2}q_a^{4}/2\epsilon ^{2})^{1/3}$ and $r_{0}=(2q_b^{2}/m \epsilon \omega _{0}^{2})^{1/3}$,
respectively, and to define the quantity $\alpha = q_b/q_a$ (with $q_a=e$ is the reference charge
taken to be equal to the elementary charge), and the screening parameter $\kappa = r_{0}/\lambda$.
In so doing, the expression of the potential energy is reduced to
\begin{equation}\label{hamiltonianII}
\begin{split}
  H &= \sum_{i}y_i^2+ \alpha \sum_{m}\sum_{n} \frac{exp(-\kappa \vert \mathbf{r}_{m} -
  \mathbf{r}_{n}\vert)}{\vert \mathbf{r}_{m} - \mathbf{r}_{n}\vert}
   \\ &+ \sum_{i>j}\frac{exp(-\kappa \vert \mathbf{r}_{i} - \mathbf{r}_{j}\vert)}{\vert \mathbf{r}_{i} - \mathbf{r}_{j}\vert} + \alpha^{2} \sum_{k>l} \frac{exp(-\kappa \vert \mathbf{r}_{k} -
\mathbf{r}_{l}\vert)}{\vert \mathbf{r}_{k} - \mathbf{r}_{l}\vert},
\end{split}
\end{equation}
and the state of the system is determined now by the ratio between charges $\alpha$, the number of
particles (which can be associated to the density), and the dimensionless screening length
$\kappa$. The temperature is expressed in units of $T_{0}=E_{0}/k_B$, where $k_B$ is the Boltzmann
constant.

The minimum energy configurations were found on the one hand by analytical calculations, and on
the other hand also through molecular dynamics simulations. In the numerical simulations we
typically considered $300$ particles, together with periodic boundary conditions in the unconfined
direction in order to mimic an infinite system.

The present model system does not address effects due to frictional dissipation. In spite of the
primary importance of friction to the motion of the particles in real systems, the ground state
configurations are not affected by it.

\section{Ground state crystal structures}
\label{sec:results}

\begin{figure}
\begin{center}
\includegraphics[scale=2.0]{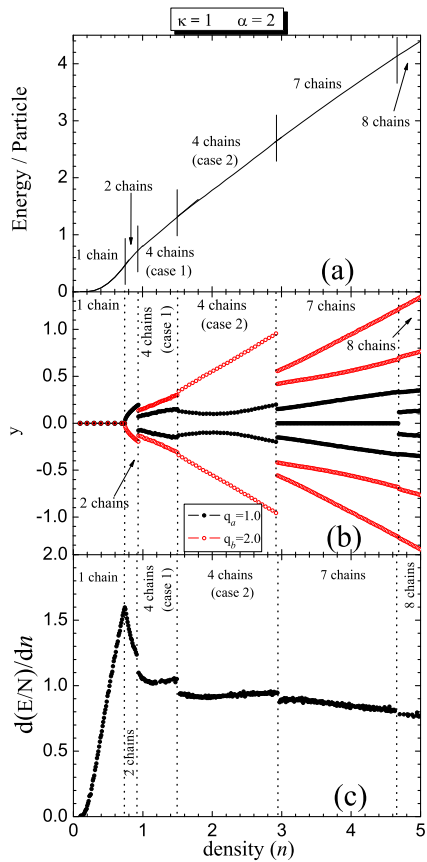}
\caption{(Color online) (a) The energy per particle, and (b) the lateral position of the chains as
a function of the density for $\kappa=1$ and $\alpha=2$. The black solid circles are the particles
with $q_a/e =1$ and the red open circles with $q_b/e=2$.  (c) The first derivative of the energy
with respect to the density.}\label{fig:ene_y_k1a2}
\end{center}
\end{figure}

In view of the large parameter space, and in order to show the increased richness of a binary
system we discuss first the case $\alpha=2$. We only considered equal mass particles. If masses
are taken different the two types of particles will feel a different confinement potential which
will result in a quantitative modification of the phase diagram \cite{wand05}.

In this section we will present the structural results for our binary Q1D system. The results were
obtained both analytically and numerically.  In the former, we calculate the energy per particle
for configurations with different number of chains, different arrangement of the particles within
the chains, and minimize such expressions with respect to the distance between chains. To study
the dynamical properties of the system we use the harmonic approximation in order to obtain the
normal modes of the ground state configuration. This is shown in section \ref{sec:phonons}.

Before starting to present our results it is convenient to introduce here a dimensionless linear
density ($n$), defined as the ratio between the total number of particles in the unit cell of the
crystalline arrangement and the length of the unit cell in the unconfined direction. This is
similar to the linear density defined previously in Ref. \onlinecite{gio04}.

First, we present the minimum energy configurations and structural transitions as a function of
the range of the interaction potential and the linear density of the system. More specifically, we
study how the ground state configurations depend on the the screening parameter ($\kappa$). Next,
we study the system as a function of the ratio between the charges of the two types of particles
($\alpha$). In the present study we limit ourselves to an equal fraction of particles with
distinct charges.

\subsection{Dependence on $\kappa$}

In general, the different charged particles organize themselves in chains. The number of chains,
and how particles are placed in them, depends on $n$ , the screening parameter ($\kappa$), and the
ratio between the charges ($\alpha$). In this sub-section, we show how the crystalline structure
of the Q1D system depends on $\kappa$ and $n$. The ratio between charges will be kept fixed
$\alpha=2.0$.

Lets first consider the case with $\kappa=1$. When $n$ increases the following sequence for the
first five ground state configurations is observed: one chain $\rightarrow$ two chains
$\rightarrow$ four chains (case 1) $\rightarrow$ four chains (case 2) $\rightarrow$ seven chains
(see Fig. \ref{fig:configs}). The energy per particle as a function of $n$ is shown in Fig.
\ref{fig:ene_y_k1a2}(a). As can be seen in Fig. \ref{fig:configs}, in the one-chain regime the
particles with distinct charges are arranged alternately, and equally spaced on the $x$-axis,
where the confinement potential is zero. In this case, the unit cell consists of two particles,
one of each type (Fig. \ref{fig:configs}). The linear density is $n = r_0/a$, and the $x$
coordinate of the particles with distinct charges are $x_i = 2ia$ and $x_i = (2i-1)a$, with $i=
\pm 1, \pm 2, \pm 3, ...$. The energy per particle for the one-chain regime is
\begin{eqnarray}\label{ene1chain}
E_{1}= (1+\alpha^{2})\frac{n}{4}\sum_{j=1}^{\infty}\frac{e^{-2\kappa j/n}}{j}+
\alpha\frac{n}{2}\sum_{j=1}^{\infty}\frac{e^{-2\kappa (j-0.5)/n}}{(j-0.5)},
\end{eqnarray}
\noindent were the first term accounts for the interaction between particles of the same type,
while the last one represents the interaction between particles with different charges.

When $n \approx 0.74155$ the two different particles start to move off chain, alternating in the
positive and negative $y$-direction, leading to a new ground state configuration consisting of two
chains. We obtained the following expression for the energy per particle,

\begin{equation}\label{ene2chain}
\begin{split}
E_{2}&=(1+\alpha^{2})\frac{n}{4}\sum_{j=1}^{\infty}\frac{e^{-2\kappa j/n}}{j} \\
&+ \alpha\frac{n}{2}\sum_{j=1}^{\infty}
\frac{exp[-2\kappa\sqrt{(j-0.5)^{2}+c^{2}}/n]}{\sqrt{(j-0.5)^{2}+c^{2}}} + \frac{c^{2}}{n^{2}},
\end{split}
\end{equation}

\noindent where $c=d/a$ is the dimensionless distance between the chains (Fig. \ref{fig:configs}).
The two-chain regime is characterized by a remarkable asymmetry in the charge distribution. It is
interesting to comment that a similar behavior was experimentally observed previously in a
two-component Coulomb crystals in a linear ($quasi$-3D) Paul traps and in a binary system of
charged dry grains \cite{hornekaer01,mehrotra}. The charge segregation characterizing the
two-chain regime is a consequence of a spontaneous symmetry breaking which occurs through a
continuous structural transition. This is made clear in Fig. \ref{fig:ene_y_k1a2}(b), where the
lateral position of the chains is plotted as a function of the density. We found that with the
exception of the transition from the one-chain to the two-chain regime, all other structural
transitions are characterized by a discontinuity in the lateral position of the chains which is
associated with a first order \lq{}\lq{}structural\rq{}\rq{} transition. In this case, the first
derivative of the energy with respect to the density is discontinuous at the transition point
[Fig. \ref{fig:ene_y_k1a2}(c)]. For the continuous transition (second order structural changes) a
discontinuity is observed only in the second derivative of the energy with respect to the density.

\begin{figure}
\begin{center}
\includegraphics[scale=0.9]{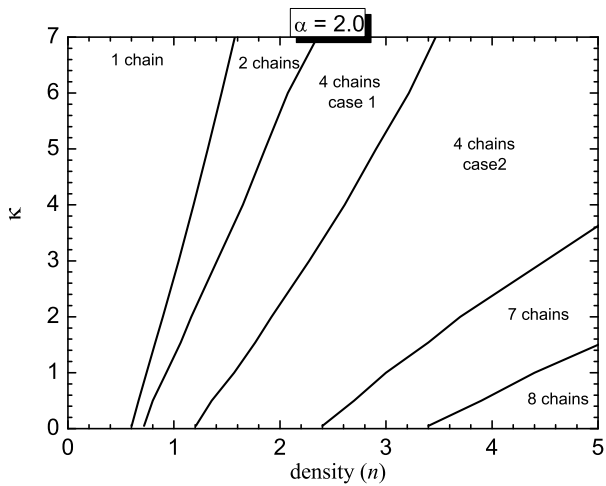}
\caption{The zero-temperature $\kappa - n$ phase diagram for $\alpha=2$. }\label{fig:phasediag_k}
\end{center}
\end{figure}

A further increase in the density brings the system to the four-chain regime. In this case, the
different types of particles are again symmetrically distributed with respect to the $x$-axis, but
still segregated. The four-chain configuration is observed in two different minimum energy
configurations, defined here as case 1 and case 2. In the four-chain (case 1) regime the rows with
the same type of charges are displaced with respect to each other by a distance $a/4$ along the
unconfined direction ($x$-axis), while neighbor rows with distinct types of particles are
displaced by a distance $a/2$ along the $x$-axis (Fig. \ref{fig:configs}). Also, the distance
between the internal chains (consisting of particles with the same charge) is larger than the
distance between the internal chains and the external ones [Fig. \ref{fig:ene_y_k1a2}(b)]. This is
interesting because the interaction between chains with distinct charges is intuitively expected
to be larger than the interaction between chains with the same lower charge. In the four-chain
(case 2) regime, chains with the same charge are displaced by $a/2$ with respect to each other
along the unconfined direction. The distance between chains has an opposite behavior to the one of
case 1, i.e. the distance between chains with distinct charges is larger than the distance between
the internal rows [Fig. \ref{fig:ene_y_k1a2}(b)].

\begin{figure}
\begin{center}
\includegraphics[scale=1.7]{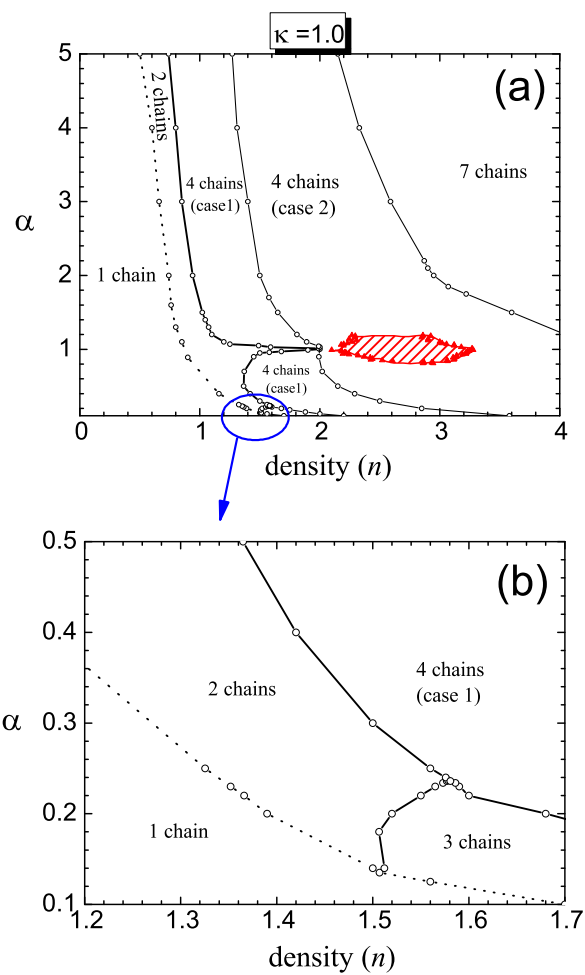}
\caption{(Color online) (a) The zero-temperature $\alpha - n$ phase diagram for $\kappa=1$. (b)
Zoom of the small $\alpha$ and intermediate density regime. Solid (dotted) lines are first
(second) order structural transitions. The symbols are the calculated points. The red dashed
region corresponds to the values of $\alpha$ and $n$ for which the two types of particles can be
found mixed in the same chain.}\label{fig:phasediag_alpha}
\end{center}
\end{figure}

The specific expressions for the energy per particle for the four-chain configurations, as well as
for other arrangements presented in Fig. \ref{fig:configs} can be obtained directly from Eq.
(\ref{hamiltonianII}). The three-chain configuration shown in Fig. \ref{fig:configs} does not
become a minimum energy arrangement for $\alpha=2$. However, as will be shown in the next section,
for some particular values of $\alpha$ such a configuration was found as the ground state.

As can be observed in Fig. \ref{fig:ene_y_k1a2}, when the density is increased the system tends to
crystallize in structures consisting of a large number of chains. In such cases it is not trivial
to predict the minimum energy arrangement of the particles. To guide our analytical analysis we
resort to numerical molecular dynamic simulations. Due to the larger number of particles
considered in the simulations, there are many stable configurations with energy only slightly
different from the minimal one. In fact, it is extremely difficult to reach the ground state
configuration, even in our numerical simulations. The final configuration typically observed in
the simulation was a mixture of several structures. We used this output as a hint to search for
the minimal energy arrangement. In Fig. \ref{fig:configs} only the stable configurations that have
minimal energy are presented, but many others were considered.

The minimum energy structures presented so far were also found when we minimized the analytical
expressions for the energy per particle with respect to the distance between chains for different
values of $\kappa$ and $n$ ($\alpha=2$). The result is presented as a zero-temperature $\kappa -
n$ phase diagram in Fig. \ref{fig:phasediag_k}. As a general feature, the interval of density in
which a particular phase is observed increases with increasing $\kappa$. Note that the larger the
value of $\kappa$ the shorter the range of inter-particle interaction potential. This means that
the critical distance between adjacent particles necessary to produce an interaction strong enough
to change the phase structure will be smaller, and consequently, the critical density will be
larger.

\subsection{Dependence on $\alpha$}

Now we will study how the crystal structure of our system depends on the density ($n$) when
different values of the ratio between charges ($\alpha$) are considered. In this section, we fixed
the screening parameter to $\kappa=1$ which is a typical value in colloidal systems and dusty
plasmas.

For different values of $\alpha$ the configurations are the same as the ones presented in Fig.
\ref{fig:configs}, but which type of particle is located in the internal and in the external
chains depends whether $\alpha$ is $<1$ or $>1$. In general, particles with smaller charge are
located in the more internal chains.

The zero temperature $\alpha - n$ phase diagram is presented in Fig. \ref{fig:phasediag_alpha}(a).
For $\alpha>1$, the same sequence of structures previously shown in Fig. \ref{fig:phasediag_k} are
found. The behavior is very different for $\alpha<1$. In this case, new structures appear or
disappear depending on the values of $\alpha$ and $n$. For example, an arrangement with three
chains (see Fig. \ref{fig:configs}) is observed for $\alpha \lesssim 0.237$, as seen in Fig.
\ref{fig:phasediag_alpha}(b). In the interval $0.136 \lesssim \alpha \lesssim 0.237$ the
three-chain regime appear as the ground state configuration for intermediate density. The
two-chain $\rightarrow$ three-chain transition is a first order structural transition.

For $\alpha \lesssim 0.136$ the two-chain regime is no longer observed as the ground state for any
value of $n$. Instead, the system changes continuously (second order structural change) from the
one- to the three-chain configuration shown in Fig. \ref{fig:configs} as the density is increased
[Fig. \ref{fig:phasediag_alpha}(b)] through a zig-zag transition. The transition three-chain
$\rightarrow$ four-chain (case 1) is a first order transition for all values of $\alpha$.

\begin{figure}
\begin{center}
\includegraphics[scale=0.8]{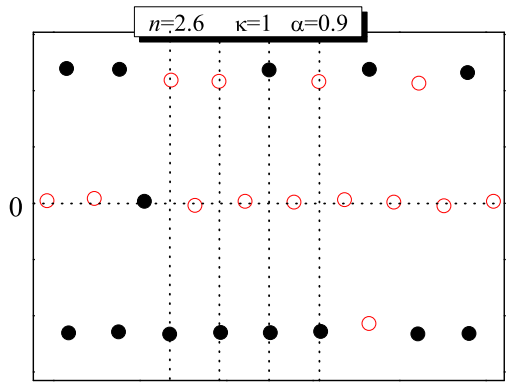}
\caption{(Color online) Example of the mixed configuration obtained from molecular dynamic
simulations. Black solid circles represent the reference charge $q_a/e=1$, while red open symbols
are the other particles with charge $q_b/e=\alpha$.}\label{fig:mixed_config}
\end{center}
\end{figure}

For $\alpha=1$ the particles are identical, and the present model is identical to the one studied
previously in Ref. \onlinecite{gio04}. We found, in agreement with Ref. \onlinecite{gio04}, that
the system presents the following sequence of five different minimum energy structures as the
density is increased: one chain $\rightarrow$ two chains $\rightarrow$ four chains $\rightarrow$
three chains $\rightarrow$ four chains $\rightarrow$ five chains. These transitions occur
respectively at densities n=0.8946, 2.0312, 2.1389, 3.3222, 4.7054. Notice that such a sequence is
different from the one found in the case $\alpha \neq 1$ [Fig. \ref{fig:phasediag_alpha}(a)]. It
should be emphasized that the three-chain structure observed when $\alpha=1$ is distinct from the
one presented in Fig. \ref{fig:configs}. In the three-chain configuration of the equal charged
particles case the rows present the same density and are displaced with respect to each other by a
distance $a/2$ along the $x$-axis. This is not the case for the three-chain arrangement commented
before when $0.136 \lesssim \alpha \lesssim 0.237$ for intermediate values of density [see Fig.
\ref{fig:configs}]. In the next section, we directed our attention to the structure of the present
binary system when the distinct charges are almost equal ($\alpha \approx 1$).

\subsection{Region around $\alpha=1$: disordered phase}

Notice that in the density region $1\lesssim n \lesssim 2$ the four-chain (case 1) phase is
strongly suppressed in favor of the two-chain phase, which for $\alpha=1$ extends now up to
$n=2.0312$. Beyond this density, there is a small density interval where the four-chain (case 2)
regime is found which for $\alpha=1$ is identical to the previously found four-chain phase
\cite{gio04}. In the density interval $2.0312 \lesssim n \lesssim 3.3222$ we found a small region
[red shaded area in Fig \ref{fig:phasediag_alpha}(a)] where the two types of particles are almost
randomly distributed over three-chains which for $\alpha \neq 1$ were found to be no longer
perfectly straight (see Fig. \ref{fig:mixed_config}).

This disordered region, where the chain-like structures are the same as the ones observed in the
case $\alpha=1$, are characterized by the presence of particles with distinct charges at the same
row. Note that the red dashed region of the $\alpha - n$ phase diagram is inside the region
associated with the four-chain (case 2) regime [Fig. \ref{fig:phasediag_alpha}(a)]. For $n \gtrsim
3.3221$ the four-chain (case 2) regime [Fig. \ref{fig:configs}] is observed for any value of
$\alpha$.

\begin{figure}
\begin{center}
\includegraphics[scale=0.8]{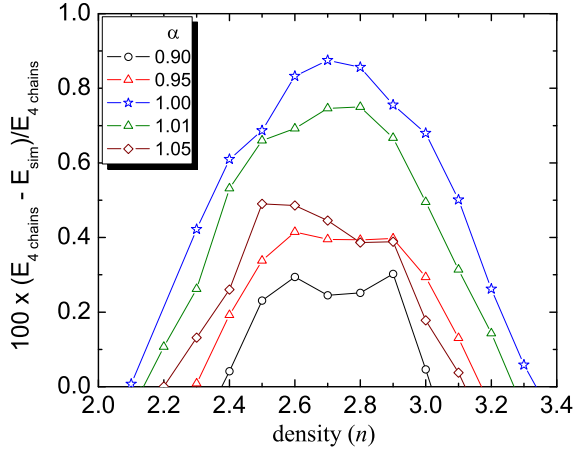}
\caption{(Color online) The relative difference between the energy of the four-chain (case 2)
regime, obtained analytically, and the energy of the mixed configuration, obtained from molecular
dynamic simulations, for values of $\alpha$ in the mixed region presented in Fig.
\ref{fig:phasediag_alpha}(a). }\label{fig:ene_mixed}
\end{center}
\end{figure}

The mixed or disordered configurations with three chains were obtained through molecular dynamic
simulations. Only in the mixed region of Fig. \ref{fig:phasediag_alpha}(a), the energy of such
configurations were observed to be smaller than those obtained analytically from the proposed
regular structures (Fig. \ref{fig:configs}). In Fig. \ref{fig:ene_mixed}, the percentage of the
difference in energy between the four-chain (case 2) regime and the mixed configurations inside
the mixed region (dashed) of Fig. \ref{fig:phasediag_alpha}(a) is presented as a function of the
density. The largest  difference in energy was observed for the case $\alpha=1$ where it was less
than 1 $\%$.

Note that asymmetric mixed configurations were found recently in a different system of binary
charged particles which were confined by a circular parabolic\cite{wand05} or hard
wall\cite{mangold04,kwinten06} potential.

\section{Phonon spectrum}
\label{sec:phonons}

\begin{figure}
\begin{center}
\includegraphics[scale=1.5]{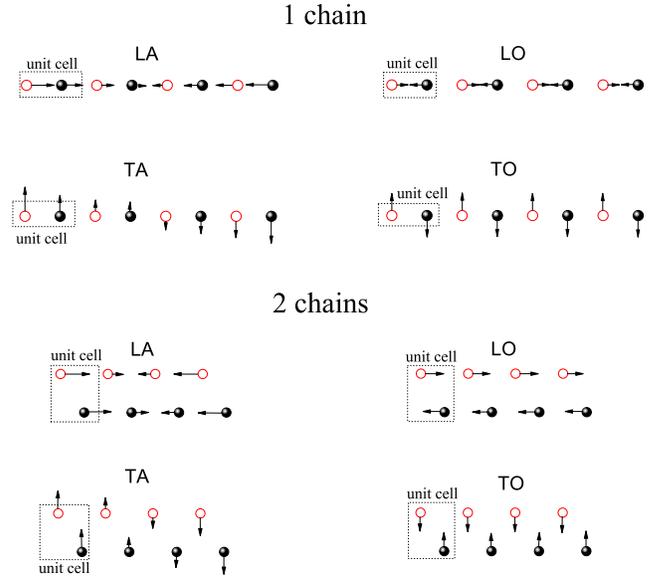}
\caption{(Color online) The motion of the particles for each type of normal modes for the one- and
two-chain regimes. }\label{fig:phonons_sketch}
\end{center}
\end{figure}

We analyze how the normal mode spectrum of the present binary system (in the absence of frictional
dissipation \cite{gio04ii}) behaves as a function of $\alpha$, $n$, and $\kappa$. Taking into
account the translational invariance of the system in the unconfined direction ($x$-axis), we
calculated the normal modes within the harmonic approximation \cite{pines}. Since we are strictly
dealing with a two-dimensional system, the number of degrees of freedom per unit cell is twice the
number of particles in the unit cell (the unit cell are indicated by dotted boxes in Fig.
\ref{fig:configs}). Therefore, if $l$ is the number of particles per unit cell there will be $2l$
branches for the phonon dispersion curves. Half of these branches correspond to oscillations along
the chain ($x$-axis - longitudinal modes), while the others are associated with vibrations along
the confinement direction ($y$-axis - transverse modes). If particles in the unit cell present an
in-phase vibration, the mode is also dominantly acoustical, while the opposite out-of-phase
oscillation determines the optical mode. In general, a normal mode can be classified in one of the
following classes: longitudinal optical (LO), longitudinal acoustical (LA), transverse optical
(TO), or transverse acoustical (TA). The motion of the particles for each type of normal mode for
the one- and two-chain regimes is illustrated in Fig. \ref{fig:phonons_sketch}.

In the harmonic approximation the normal modes are obtained by solving the system of equations

\begin{equation}
(\omega^2 \delta_{\mu\nu,ij} - D_{\mu\nu,ij})Q_{\nu,j}=0,
\end{equation}

\noindent where $Q_{\nu,j}$ is the displacement of particle $j$ from its equilibrium position in
the $\nu$ direction, $\mu$ and $\nu$ refer to the spatial coordinates $x$ and $y$,
$\delta_{\mu\nu,ij}$ is the unit matrix and $D_{\mu\nu,ij}$ is the dynamical matrix, defined by

\begin{equation}
D_{\mu\nu,ij}=\frac{1}{m}\sum_{u}\phi_{\mu,\nu}(u) e^{-iuqa},
\end{equation}

\noindent where $u$ is an integer assigned to each unit cell. The force constants are given by

\begin{equation}
\phi_{\mu,\nu}(u)=\partial_\mu \partial_\nu \frac{\exp(-\kappa \sqrt{(x-x')^2 +
(y-y')^2})}{\sqrt{(x-x')^2 + (y-y')^2}},
\end{equation}

\noindent with $(x-x') \in [{au,a(u-1/2)}]$, and $(y-y')=$ interchain distance with $(x,y)$ and
$(x',y')$ the equilibrium positions of the particles in the unit cell, and

\begin{equation}
\phi_{\mu,\nu}(u=0)=-\sum_{u\neq0}\phi_{\mu,\nu}(u).
\end{equation}

\noindent The phonon frequency is given in units of $\omega_0 / \sqrt{2}$. As an example, the
complete dynamical matrix for the one-chain regime is given in Appendix \ref{app:appendixB} as an
example.

The frequencies for the one-chain configuration are given by
%\begin{equation}
%\omega_l =\sqrt{(A_1+A_5) \pm \sqrt{(A_1+A_5)^2-4(A_1A_5-A_3^2)}}
%\end{equation}
\begin{equation}
\omega_l =\sqrt{(A_1+A_3) \pm \sqrt{(A_1-A_3)^2+4A_5^2}}
\end{equation}
for the longitudinal modes, and by
%\begin{equation}
%\omega_t =\sqrt{1+(A_2+A_6) \pm \sqrt{(A_2+A_6)^2-4(A_2A_6-A_4^2)}}
%\end{equation}
\begin{equation}
\omega_t =\sqrt{1+(A_2+A_4) \pm \sqrt{(A_2-A_4)^2+4A_6^2}}
\end{equation}
for the transverse modes. The parameters $A_1,A_2,...,A_6$ are given in Appendix
\ref{app:appendixB}. The wave number $k$ for the one- and the two-chains regimes is in units of
$\pi/2a$, where $2a$ is the length of the unit cell in the $x$-direction.

\begin{figure}
\begin{center}
\includegraphics[scale=0.9]{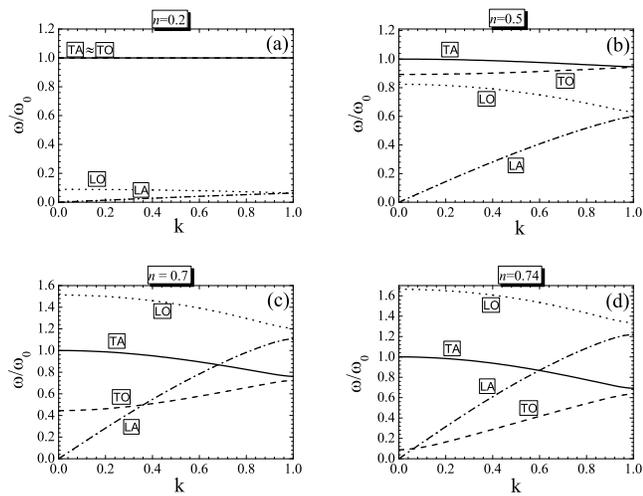}
\caption{The phonon spectrum for the one chain configuration with $\kappa=1$ and $\alpha=2$ for
different values of the density. The transverse acoustical (TA), the transverse optical (TO), the
longitudinal optical (LO), and the longitudinal acoustical (LA) phonon dispersion curves are
identified, respectively, by the solid, dashed, dotted and dot-dashed curves.
}\label{fig:phonons_k1a2}
\end{center}
\end{figure}

In Fig. \ref{fig:phonons_k1a2} the phonon spectrum of the system in the one-chain regime is
presented for different values of the density $n$, and fixed values of $\kappa=1$ and $\alpha=2$.
As can be observed, there is a clear dependence of the dispersion curves on $n$. For small values
of the density, e.g. $n=0.2$, the transverse modes (optical and acoustical) are almost the same
[Fig. \ref{fig:phonons_k1a2}(a)]. As the density increases the frequency of the TO mode decreases.
An interesting feature to keep in mind is that specially for the TO mode, of the one-chain regime,
there is a resulting force in the direction of the confinement potential ($y$-direction) as a
consequence of the repulsive interaction between particles with distinct charges. For a
sufficiently small density, the distance between adjacent particles becomes large enough to make
the interaction between particles irrelevant to affect the oscillation of such particles around
the equilibrium positions. In this case, the repulsive force in the $y$-direction is very small
and the TO mode will be mainly determined by the confinement potential. A similar explanation can
be given for the behavior of the TA mode, but in this case the repulsive force in the
$y$-direction is smaller than that for the TO mode. Therefore, for a sufficiently small density
the transverse modes will be determined predominantly by the restoring force due to the
confinement potential, which is the reason why such modes become almost indistinguishable for
small densities.

When the density is increased the distance between adjacent particles becomes smaller, the
repulsive force between them increases and acts as a retarding force. As a consequence, the
frequency of the TO oscillations decreases.

\begin{figure}
\begin{center}
\includegraphics[scale=0.9]{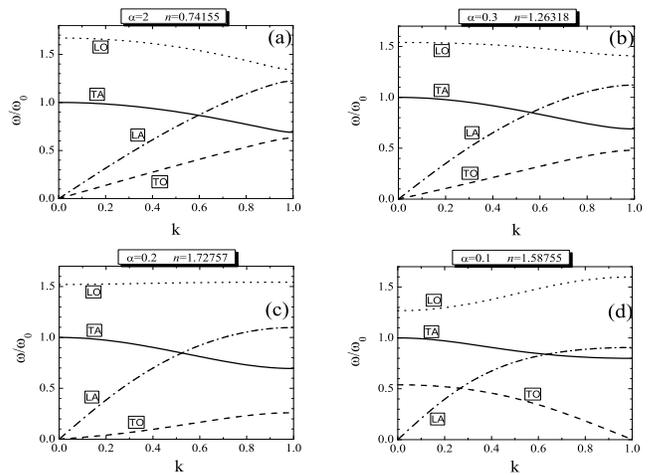}
\caption{The normal modes of the system at the transition (a), (b), (c) from the one-chain to the
two-chain regime, and (d) from the one-chain to the three-chain structure. The TA, TO, LO, and LA
relation dispersion curves are given, respectively, by the solid, dashed, dotted, and dot-dashed
curves. }\label{fig:phonons_k1_transit}
\end{center}
\end{figure}

The LO mode has the opposite behavior as compared to the TO mode, i.e. there is a hardening of
such mode when the density increases. This is a consequence of the larger repulsion due to the
closer proximity between particles. A hardening with increasing density is also observed for the
LA mode.

For the LA mode, a linear dispersion curve as $k \rightarrow 0$ is found. This means that in this
limit the longitudinal wave propagates with constant velocity along the chain, and such a velocity
increases with increasing density (Fig. \ref{fig:phonons_k1a2}).

As can be observed in Fig. \ref{fig:phonons_k1a2}, specially in the limit $k \rightarrow 0$, the
TO branch decreases with increasing density. For $n \approx 0.74155$ the system with $\kappa=1$
and $\alpha=2$ undergoes a second order structural phase transition to the two-chain regime. From
Fig. \ref{fig:phonons_k1_transit}(a) we notice that the TO mode is zero for $k=0$ and becomes
linear in the limit $k \rightarrow 0$, similar to the LA mode. Such a behavior characterizes a
continuous structural transition.

\begin{figure}
\begin{center}
\includegraphics[scale=1.7]{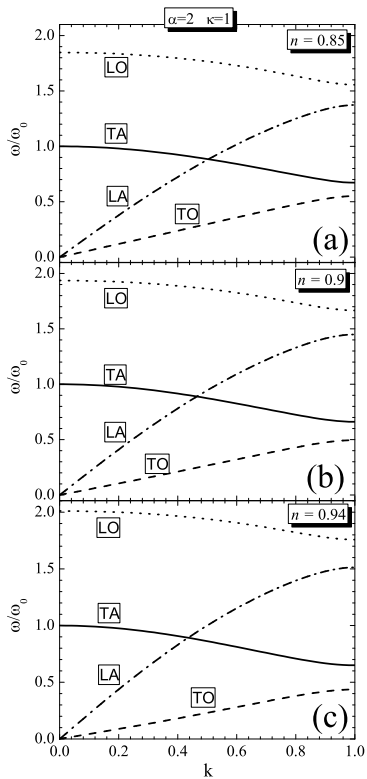}
\caption{The phonon spectrum for the two-chain configuration for different values of the density
and with $\kappa=1$ and $\alpha=2$. The transverse acoustical (TA), the transverse optical (TO),
the longitudinal optical (LO), and the longitudinal acoustical (LA) phonon dispersion curves are
identified, respectively, by the solid, dashed, dotted, and dot-dashed curves.
}\label{fig:phonons_2ch_k1a2}
\end{center}
\end{figure}

In Fig. \ref{fig:phonons_k1_transit} the dispersion curves for systems with $\kappa=1$ and
distinct values of $\alpha$ are presented. In each case, we consider the critical density at which
the system transits from the one-chain to the two-chain [Figs. \ref{fig:phonons_k1_transit}(a),
\ref{fig:phonons_k1_transit}(b), \ref{fig:phonons_k1_transit}(c)] or to the three-chain regime
[Figs. \ref{fig:phonons_k1_transit}(d)]. As can be observed in Figs.
\ref{fig:phonons_k1_transit}(a), \ref{fig:phonons_k1_transit}(b), \ref{fig:phonons_k1_transit}(c)
for $\alpha=2$, $\alpha=0.3$, $\alpha=0.2$, respectively, the TO mode is zero for $k=0$ and is
linear for small $k$-values.

In Fig. \ref{fig:phonons_k1_transit}(d), $\alpha=0.1$, the dispersion curve for the TO mode is
qualitatively different from the previous ones. The softening of the TO mode is still observed,
but now at the edge of the first Brilloum zone ($k=1$). In this case, the system changes directly
from the one-chain to the three-chain regime.

The dependence of the phonon dispersion curves on the linear density for the two-chain
configuration, in the case with $\kappa=1$ and $\alpha=2$, is presented in Fig.
\ref{fig:phonons_2ch_k1a2}. In general, the frequencies of the longitudinal (transverse) branches
increases (decreases) with increasing density. The explanation for such a behavior is similar to
the one presented previously for the one-chain regime.

\begin{figure}
\begin{center}
\includegraphics[scale=0.7]{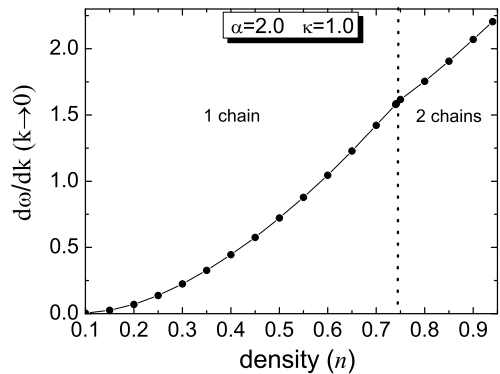}
\caption{The sound velocity obtained from the LA phonon mode for the one and two-chain
configurations as a function of density for $\kappa=1$ and $\alpha=2$.}\label{fig:sound_velocity}
\end{center}
\end{figure}

The sound velocity, i. e. $v_g = d \omega/dk\mid_{k=0}$, corresponding to the LA phonon mode is
shown in Fig. \ref{fig:sound_velocity} as a function of the density for $\alpha=2$. Notice that
the sound velocity increases with density as expected. At $n=0.74155$ a second order structural
phase transition occurs to a two-chain configuration. At this phase transition the monotonic
behavior of $v_g$ is changed because of the decrease in the density of particle along each chain
and the continuous increasing of the separation between the chains.

The general qualitative behavior of the phonon dispersion curves presented so far for $\kappa=1$
is also found for different values of the screening parameter.

\section{Conclusions}
\label{sec:conclusions}

We studied a $quasi$-one-dimensional binary system of charged particles confined in a parabolic
trap. The general structure and the main features of the normal mode spectrum were analyzed. The
structural and dynamical properties were studied as a function of the density, the range of the
interacting potential, which is an experimental tunable parameter in systems like colloidal
dispersions, dusty plasmas or even in a binary system of hard-spheres, and the parameter
characterizing the binary system, namely, the ratio between charges of the two types of particles.
We found a very rich variety of ground state configurations, some of them are even not symmetric
around the symmetry axis (i.e. $y=0$) of the 1D confinement potential.

The number of chains as well as the internal structure in such chains are a function of the
parameters of the system ($\alpha$, $n$, and $\kappa$). The set of structures was summarized in
zero temperature phase diagrams which relate the ratio between the distinct types of particles and
the screening parameter of the interaction potential with the linear density of the system. The
structural transitions between the distinct phases were characterized as being of first or second
order. A spontaneous symmetry breaking in the charge distribution of the system was found
corresponding to a structural transition that is characterized by a continuous change in the
lateral position of the particles. A disordered phase was found in a small region of the
four-chains (case 2) part of the phase diagram where the particles are $quasi-$randomly
distributed over three chains. The disordered three-chain configuration transits in a
discontinuous way to the four-chain (case 2) ordered phase.

The phonon dispersion curves for the one- and two-chain structures were obtained and analyzed for different
values of $\alpha$ and $n$. In general, the normal mode frequencies depend on the linear density
of the system. The transverse optical mode decreases monotonically with increasing density, and
such a feature was observed for all values of $\kappa$ and $\alpha$. The opposite behavior was
observed for the longitudinal modes (optical and acoustical), i.e. there is a hardening of such
modes with increasing density.

A remarkable softening of one of the branches of the dispersion curves was found at the second
order structural transition. The softening of the TO phonon dispersion curve may occur for at the
center or at the edge of the first Brillouin zone ($k=\pi/2a$), depending, respectively, if the
transition is between the one- and the two-chain configuration or between the one- and the
three-chain structure.

\section{Acknowledgments}

Fruitful discussions and collaboration with H.~A.~Carmona are gratefully acknowledged. WPF, PWSO,
JCNC and GAF were supported by the Brazilian National Research Councils: CNPq and CAPES and the
Ministry of Planning (FINEP). FMP was supported by the Flemish Science Foundation (FWO-Vl).

\begin{widetext}
\appendix

\section{}
\label{app:appendixB}

The matrix $\omega^2 \textbf{I} - \textbf{D}$ (where $\textbf{I}$ is the unit matrix and
$\textbf{D}$ is the dynamical matrix) is used in the calculation of the normal modes for the one-
and two-chains configurations. The dynamical matrix is of the form

$$
\left[\begin{array}
{rrrr}
\omega^{2}-A_{1}&  0& -A_{5}&0\\
0&   \Delta\omega^{2}-A_{2} &0&-A_{6}\\
-A_{5}&0&\omega^{2}-A_{3}&0\\
0&-A_{6}&0&     \Delta\omega^{2}-A_{4}\\
\end{array}\right]\label{matrix} ,
$$

\noindent where $\Delta\omega^{2} = \omega^2 -\omega_{0}^{2}$. The quantities $A_1$, $A_2$ ($A_3$,
$A_4$) are associated with the interaction between particles with charge $q_a$ ($q_b$), while
$A_5$ and $A_6$ account for the interaction between distinct charges. For the one-chain
configuration, such parameters are given by

\begin{eqnarray}
\begin{aligned}
A_1&=\alpha\sum_{j=1}^{\infty}n^{3}\frac{e^{-\kappa(2j-1)/n}}{(2j-1)^{3}}\Big[2+\frac{2\kappa
(2j-1)}{n}+\frac{\kappa^2(2j-1)^{2}}{n^{2}}\Big] \hfill
\\& +\sum_{j=1}^{\infty}n^{3}\frac{e^{-2\kappa j/n}}{(2j)^{3}}\Big[2+\frac{4\kappa
j}{n}+\frac{(2\kappa j)^{2}}{n^{2}}\Big][1-cos(2kja)],
\end{aligned}
\end{eqnarray}

\begin{eqnarray}
\begin{aligned}
A_2&=-\alpha\sum_{j=1}^{\infty}n^{3}\frac{e^{-\kappa(2j-1)/n}}{(2j-1)^{3}}\Big[1+\frac{\kappa
(2j-1)}{n}\Big] \\ &- \hfill\sum_{j=1}^{\infty}n^{3}\frac{e^{-2\kappa
j/n}}{(2j)^{3}}\Big[1+\frac{2\kappa j}{n}\Big][1-cos(2kja)],
\end{aligned}
\end{eqnarray}

\begin{eqnarray}
\begin{aligned}
A_3&=\alpha\sum_{j=1}^{\infty}n^{3}\frac{e^{-\kappa(2j-1)/n}}{(2j-1)^{3}}\Big[2+\frac{2\kappa
(2j-1)}{n}+\frac{\kappa^2(2j-1)^{2}}{n^{2}}\Big] \hfill
\\ &+\alpha^{2}\sum_{j=1}^{\infty}n^{3}\frac{e^{-2\kappa j/n}}{(2j)^{3}}\Big[2+\frac{4\kappa
j}{n}+\frac{(2\kappa j)^{2}}{n^{2}}\Big][1-cos(2kja)],
\end{aligned}
\end{eqnarray}

\begin{eqnarray}
\begin{aligned}
A_4&=-\alpha\sum_{j=1}^{\infty}n^{3}\frac{e^{-\kappa(2j-1)/n}}{(2j-1)^{3}}\Big[1+\frac{\kappa
(2j-1)}{n}\Big] \\ &- \hfill\alpha^{2}\sum_{j=1}^{\infty}n^{3}\frac{e^{-2\kappa
j/n}}{(2j)^{3}}\Big[1+\frac{2\kappa j}{n}\Big][1-cos(2kja)],
\end{aligned}
\end{eqnarray}

\begin{eqnarray}
\begin{aligned}
A_5&=-\alpha\sum_{j=1}^{\infty}n^{3}\frac{e^{-\kappa(2j-1)/n}}{(2j-1)^{3}}\Big[2+\frac{2\kappa
(2j-1)}{n}+\frac{\kappa^2(2j-1)^{2}}{n^{2}}\Big][cos(k(2j-1)a)],
\end{aligned}
\end{eqnarray}

\begin{eqnarray}
\begin{aligned}
A_6&=\alpha\sum_{j=1}^{\infty}n^{3}\frac{e^{-\kappa(2j-1)/n}}{(2j-1)^{3}}\Big[1+\frac{\kappa
(2j-1)}{n}\Big][cos(k(2j-1)a)].
\end{aligned}
\end{eqnarray}

The dimensionless wave number $k$ is in units of $\pi/2a$.

For the two-chain regime the parameters in the dynamical matrix are of the form:

\begin{eqnarray}
\begin{aligned}
A_1&=\alpha\sum_{j=1}^{\infty}n^{3}\frac{e^{-\kappa
r/n}}{r^{3}}\Big[(2j-1)^2\Big(\frac{3}{r^2}+\frac{3\kappa
}{nr}+\frac{\kappa^2}{n^{2}}\Big)-(1+\frac{\kappa r}{n})\Big] \hfill
\\ &+\sum_{j=1}^{\infty}n^{3}\frac{e^{-2\kappa j/n}}{(2j)^{3}}\Big[2+\frac{4\kappa
j}{n}+\frac{(2\kappa j)^{2}}{n^{2}}\Big][1-cos(2kja)],
\end{aligned}
\end{eqnarray}

\begin{eqnarray}
\begin{aligned}
A_2&=\alpha\sum_{j=1}^{\infty}n^{3}\frac{e^{-\kappa
r/n}}{r^{3}}\Big[\frac{3c^2}{r^2}+\frac{\kappa^2 c^2}{n^2}+\frac{3\kappa
c^{2}}{nr}-(1+\frac{\kappa r}{n})\Big] \\& - \sum_{j=1}^{\infty}n^{3}\frac{e^{-2\kappa
j/n}}{(2j)^{3}}\Big[1+\frac{2\kappa j}{n}\Big][1-cos(2kja)],
\end{aligned}
\end{eqnarray}

\begin{eqnarray}
\begin{aligned}
A_3&=\alpha\sum_{j=1}^{\infty}n^{3}\frac{e^{-\kappa
r/n}}{r^{3}}\Big[(2j-1)^2\Big(\frac{3}{r^2}+\frac{3\kappa
}{nr}+\frac{\kappa^2}{n^{2}}\Big)-(1+\frac{\kappa r}{n})\Big] \hfill
\\ &+\alpha^{2}\sum_{j=1}^{\infty}n^{3}\frac{e^{-2\kappa j/n}}{(2j)^{3}}\Big[2+\frac{4\kappa
j}{n}+\frac{(2\kappa j)^{2}}{n^{2}}\Big][1-cos(2kja)],
\end{aligned}
\end{eqnarray}

\begin{eqnarray}
\begin{aligned}
A_4&=\alpha\sum_{j=1}^{\infty}n^{3}\frac{e^{-\kappa
r/n}}{r^{3}}\Big[\frac{3c^2}{r^2}+\frac{\kappa^2 c^2}{n^2}+\frac{3\kappa
c^{2}}{nr}-(1+\frac{\kappa r}{n})\Big] \\ &-
\hfill\alpha^{2}\sum_{j=1}^{\infty}n^{3}\frac{e^{-2\kappa j/n}}{(2j)^{3}}\Big[1+\frac{2\kappa
j}{n}\Big][1-cos(2kja)],
\end{aligned}
\end{eqnarray}

\begin{eqnarray}
\begin{aligned}
A_5&=-\alpha\sum_{j=1}^{\infty}n^{3}\frac{e^{-\kappa
r/n}}{r^{3}}\Big[(2j-1)^2\Big(\frac{3}{r^2}+\frac{3\kappa
}{nr}+\frac{\kappa^2}{n^{2}}\Big)-(1+\frac{\kappa r}{n})\Big][cos(k(2j-1)a)],
\end{aligned}
\end{eqnarray}

\begin{eqnarray}
\begin{aligned}
A_6&=-\alpha\sum_{j=1}^{\infty}n^{3}\frac{e^{-\kappa
r/n}}{r^{3}}\Big[\frac{3c^2}{r^2}+\frac{\kappa^2 c^2}{n^2}+\frac{3\kappa
c^{2}}{nr}-(1+\frac{\kappa r}{n})\Big][cos(k(2j-1)a)],
\end{aligned}
\end{eqnarray}

\noindent where $r=\sqrt{(2j-1)^2+c^2}$, and $c=d/a$ ($d$ and $a$ are indicated in Fig.
\ref{fig:configs}) is chosen in order that the eigenvalues of the dynamical matrix be positive.

\end{widetext}

\end{document}